\documentclass[12pt]{article}

\usepackage{hyperref}
\usepackage{graphicx}
\usepackage{amsmath}
\usepackage{lmodern}
\usepackage[left=0.60in,right=0.60in,top=0.60in,bottom=0.60in]{geometry}
\usepackage{enumitem}
\usepackage{pdflscape}

\title{\LARGE\textbf{Predicting the Accuracy of Early-est Earthquake Magnitude Estimates with an LSTM Neural Network: A Preliminary Analysis}}

\author{Massimo Nazaria}

\date{
    Istituto Nazionale di Geofisica e Vulcanologia (INGV)\\
    Via di Vigna Murata 605, 00143 Roma, Italy\\[2ex]
    February 26, 2021
}

\begin{document}

\maketitle

\begin{abstract}
\noindent
This report presents a preliminary analysis of an LSTM neural network designed to predict the accuracy of magnitude estimates computed by Early-est during the first minutes after an earthquake occurs. Data and code are available at \url{https://github.com/massimo-nazaria/lstm}.
\end{abstract}

\section{Introduction}

\href{http://early-est.rm.ingv.it}{Early-est} is a software system that locates earthquakes on a global scale and computes their characterization parameters by using the data provided by seismic stations in real time \cite{lomax2009}\cite{bernardi2015}. Its results are the input for an automatic procedure that first inspects the computed earthquake parameters, then evaluates the possibility that a tsunami was generated, and finally prepares the text of the first message to be issued by the tsunami warning system managed by \href{https://www.ingv.it/cat/en/the-italian-alert-system/the-tsunami-alert-centre-cat-ingv}{CAT-INGV}.
The position and size of the earthquake are recalculated at one-minute intervals, which makes them more and more accurate as seismic waves reach stations further away from the hypocenter and thus additional data becomes available. The ability to predict the accuracy of magnitude estimates during the first minutes since the occurrence of an earthquake would enable us to obtain the final magnitude in advance. While it is impractical to involve traditional programming to develop a software tool that computes such predictions, machine learning presents us with the possibility to attempt modeling a neural network which independently learns how to predict the accuracy of estimates by examining historical data without explicit instructions by the programmer. Unfortunately, we cannot assume that machine learning is a viable approach to our problem, as the data at our disposal may be unsuitable or insufficient for the automatic training of a neural network. Furthermore, to the best of our knowledge there are no neural network models in the literature we can adapt to our problem (see, e.g.,~\cite{kong2018} and references therein). In this report we present a preliminary analysis of an LSTM neural network that attempts to predict the accuracy of magnitude estimates computed by Early-est during the first minutes after the occurrence of an earthquake.

\vfill
\noindent
\makebox[0.60in][l]{This work is licensed under a \href{https://creativecommons.org/licenses/by/4.0/}{Creative Commons Attribution 4.0 International License}.}\hfill\href{https://creativecommons.org/licenses/by/4.0/}{\includegraphics[width=0.14\textwidth]{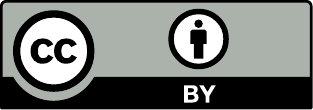}}

\clearpage

\subsection{Recurrent Neural Networks and LSTMs}

Recurrent Neural Networks (RNNs) are designed to selectively recognize patterns within input data which are organized as time series by inferring dependencies among subsequent data timesteps. RNNs are able to accomplish this with the use of a sort of internal memory which gets updated at each timestep being processed in order to keep track of meaningful information over the entire sequence of input. Nevertheless, the fact that RNNs repeatedly update their internal memory makes these networks suffer from the well known vanishing/exploding gradient problems \cite{hochreiter1991}\cite{hochreiter1998}.
In short, RNNs exhibit the tendency to overwrite their internal memory thus losing the ability to remember useful information from past data timesteps. On the other hand, Long Short-Term Memory (LSTM) networks are a kind of RNN which deals with the aforementioned problems with the use of a more complex internal architecture, that makes the network able to decide whether information at each timestep are going to be remembered or forgotten \cite{hochreiter1997}.
This enables LSTMs to have greater control over their internal memory updates in order to effectively infer dependencies among consecutive data timesteps thus providing better results.
The fact that Early-est historical data consist of sequences of parameters computed at consecutive minutes makes LSTMs suitable for our needs. Namely, our LSTM should be able to learn how to predict the accuracy of magnitude estimates at a given minute on the basis of earthquake parameters computed at previous minutes.

\begin{figure}[ht!]
\begin{center}
\includegraphics[width=\textwidth]{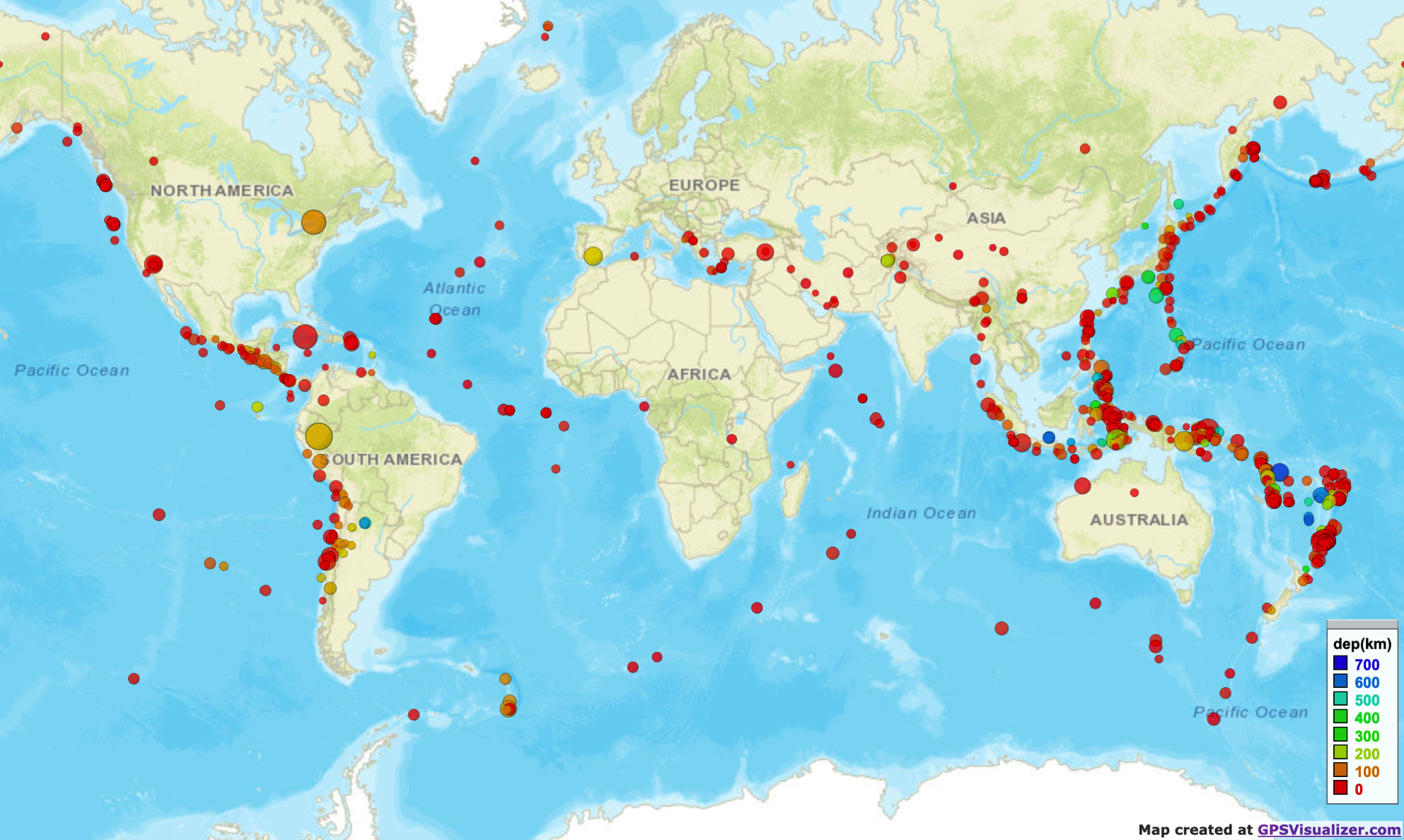}
\caption{\label{figure:dataset}Location of the 608 earthquakes used in this work. The size of the circles is proportional to the magnitude of the earthquakes; the color depends on the depth of the hypocenter.}
\end{center}
\end{figure}

\clearpage

\renewcommand\arraystretch{2.2}

\begin{table}[ht]
\centering
\begin{tabular}{|p{.25\textwidth}|p{.15\textwidth}|p{.5\textwidth}|}
\hline
\hline
\textbf{Name} & \textbf{Type} & \textbf{Description}\\
\hline
\hline
\texttt{loc\_err\_h} & Double & Ellipsoid horizontal error\\
\hline
\texttt{loc\_err\_z} & Double & Ellipsoid vertical error\\
\hline
\texttt{loc\_err\_resid} & Double & RMS of the residuals\\
\hline
\texttt{loc\_num\_st} & Integer & Number of stations used for location\\
\hline
\texttt{loc\_min\_dist} & Double & Distance of the nearest station\\
\hline
\texttt{loc\_avg\_dist} & Double & Average distance of stations\\
\hline
\texttt{loc\_std\_dist} & Double & Standard deviation distance of stations\\
\hline
\texttt{loc\_1st\_azim\_gap} & Double & Primary azimuthal gap\\
\hline
\texttt{loc\_2nd\_azim\_gap} & Double & Secondary azimuthal gap\\
\hline
\texttt{mag} & Double & Estimated magnitude of type Mwp\\
\hline
\texttt{mag\_err} & Double & Magnitude error\\
\hline
\texttt{mag\_num\_st} & Integer & Number of stations used for \texttt{mag} ($\le$ \texttt{loc\_num\_st})\\
\hline
\hline
\end{tabular}
\caption{\label{table:parameters}Early-est earthquake parameters used in this work.}
\end{table}

\section{Data Preparation}

In the database with historical data by Early-est, the 12 earthquake parameters used in this work, which are described in Table~\ref{table:parameters}, are associated with both the event IDs they refer to as well as the minutes they were computed at. Minutes are represented as integers and are relative to the time of arrival of the very first seismic data of their corresponding events.
These 12 parameters are intended to provide us with useful information about the accuracy of both location and magnitude estimates (e.g. RMS of the residuals, number of stations used) as well as the geometry of those stations where the seismic data came from (e.g. average distance of stations, primary and secondary azimuthal gap).
Studies showed how these parameters are correlated to the accuracy of the final estimates of both location and magnitude, and how more accurate locations result into more accurate magnitude estimates (see, e.g.,~\cite{bondar2004} and citations therein).

\clearpage

\renewcommand\arraystretch{2.2}

\begin{table}[h!]
\centering
\begin{tabular}{|p{.45\textwidth}|p{.45\textwidth}|}
\hline
\hline
\textbf{Early-est Version} & 1.2.4\\
\hline
\textbf{Number of Events} & 608\\
\hline
\textbf{Time Interval of the Events} & 2019/04/15--2020/01/30\\
\hline
\textbf{Minutes per Event} & 1, 2, 3, $\ldots$, 16\\
\hline
\textbf{Magnitudes at Minutes 16} & $\le$ 5\\
\hline
\textbf{Range of Magnitudes} & $[4.3, 7.8]$\\
\hline
\textbf{NULL Values} & None\\
\hline
\hline
\end{tabular}
\caption{\label{table:dataset}Information about the initial dataset.}
\end{table}

\subsection{Initial Dataset}

Our initial dataset consists of a selection of historical events, which are summarized in Table~\ref{table:dataset}. The data were computed by Early-est version 1.2.4 and comprise the 608 events in Figure~\ref{figure:dataset} located all over the world in the period of time between 15 April 2019 and 30 January 2020. For each event in the dataset, there are the values of the 12 parameters in Table~\ref{table:parameters} that were computed at minutes from 1 to 16. The magnitude estimates at minutes 16 are all greater than or equal to 5. Finally, the range of the magnitudes is $[4.3, 7.8]$ and no NULL parameter values are present.

\subsection{Definition of Accuracy}

Following a preliminary analysis of the data, we select the magnitude estimates computed at minutes 16 as the most accurate ones and define the accuracy of a given estimate $mag_t^e$ computed at the $t$-th minute of an event $e$ with the following formula:
$$accuracy(mag_t^e):=\frac{mag_t^e-mag_{16}^e+1}{2}.$$

\noindent
The accuracy of $mag_t^e$ is the difference $mag_t^e-mag_{16}^e$ between the magnitudes estimated at the minutes $t$ and 16 of the same event $e$. In order to make sure that all the accuracies fall into the range $[0, 1]$, we add 1 to the difference and divide the result by 2.\\

\begin{landscape}
\begin{figure}[ht]
\begin{center}
\includegraphics[width=1.3\textwidth]{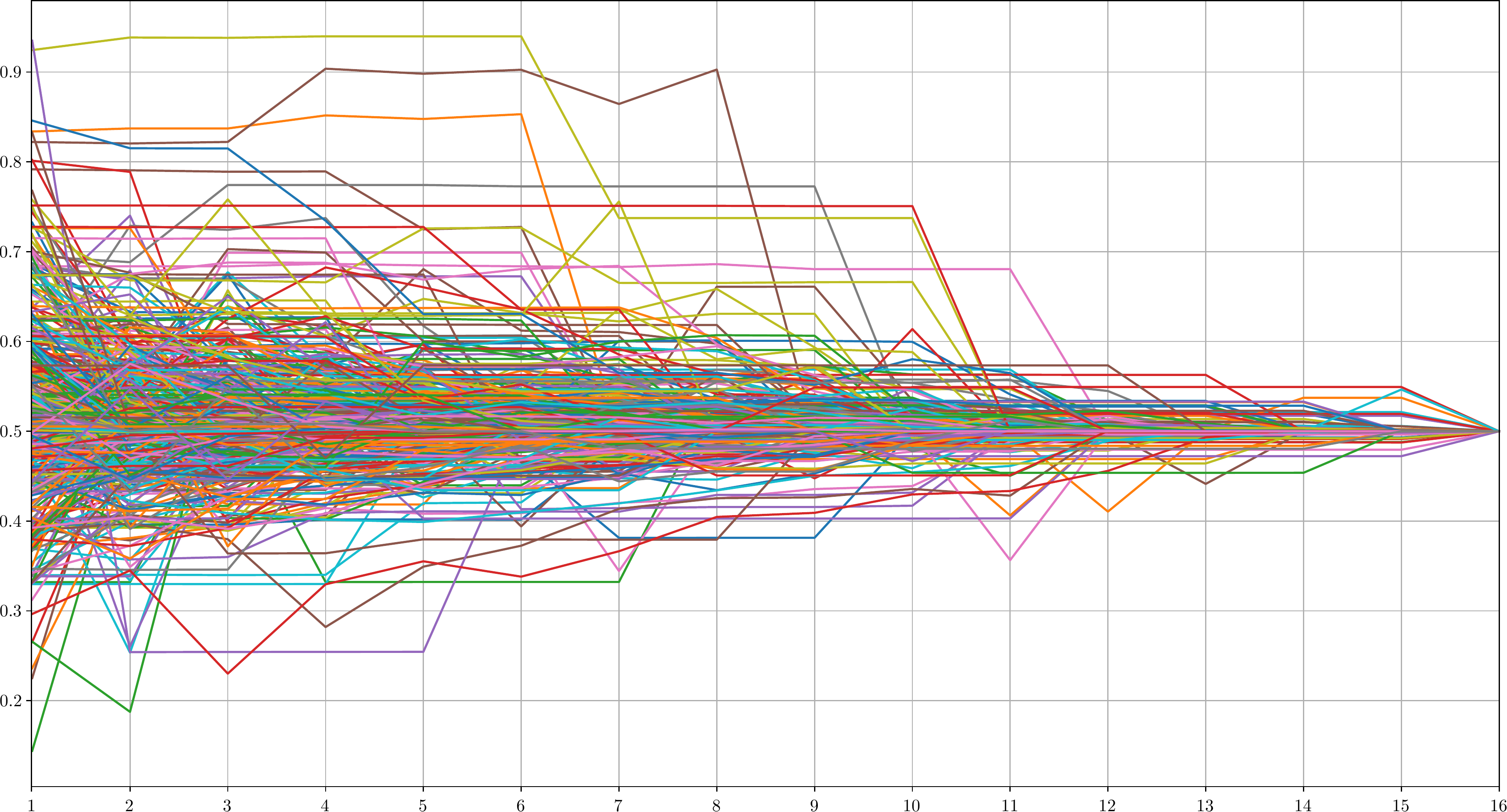}
\caption{\label{figure:accuracy}Magnitude accuracies of the 608 selected events at minutes from 1 to 16.}
\end{center}
\end{figure}
\end{landscape}

\noindent
Figure~\ref{figure:accuracy} shows the magnitude accuracies of all the 608 events at minutes from 1 to 16. The increasing availability of data used by Early-est to recompute its estimates makes the latter more and more accurate as the minutes go by.

\clearpage

\section{Feature Extraction and Labeling}
\label{section:samples}

The training samples for our neural network consist of pairs with inputs and desired outputs we refer to as features and labels, respectively. Firstly, let us define $Par_t^e(p)$ as the value of the parameter $p$ of a given event $e$ at minute $t$. Secondly, let $Par_t^e:=(Par_t^e(\texttt{loc\_err\_h}),\ldots,Par_t^e(\texttt{mag}),\ldots)$ be defined as the sequence that contains values of all the 12 parameters seen in Table~\ref{table:parameters} of a given event $e$ at minute $t$. Finally, let us define $Par^e:=(Par_1^e,Par_2^e,\ldots,Par_N^e)$ as the sequence of $N$ sequences $Par_i^e$, with $1 \le i \le N$, and $N := 16$.

\subsection{Features}

For each event $e$, we prepare input features from $Par^e$ by computing a number of $N-K+1$ subsequences of length $K$, with $K := 2$, in this way:

\begin{itemize}[label={}]
    \item $X_1^e:=(Par_1^e, Par_2^e, \ldots, Par_K^e)$,
    \item $X_2^e:=(Par_2^e, \ldots, Par_K^e, Par_{K+1}^e)$,
    \item $\vdots$
    \item $X_{N-K+1}^e:=(Par_{N-K+1}^e, Par_{N-K+2}^e, \ldots, Par_N^e)$.
\end{itemize}

\subsection{Labels}

For each feature $X_i^e$, we prepare the corresponding label $Y_i^e$, with $1 \le i \le N-K+1$, as follows:

\begin{itemize}[label={}]
    \item $Y_1^e:=\frac{Par_K^e(\texttt{mag})-Par_N^e(\texttt{mag})+1}{2}$,
    \item $Y_2^e:=\frac{Par_{K+1}^e(\texttt{mag})-Par_N^e(\texttt{mag})+1}{2}$,
    \item $\vdots$
    \item $Y_{N-K+1}^e:=\frac{Par_{N}^e(\texttt{mag})-Par_N^e(\texttt{mag})+1}{2} = \frac{0+1}{2} = 0.5$.
\end{itemize}

\noindent
As a result, we obtain the following 15 samples $(X_i^e, Y_i^e)$ for each event $e$, with $1 \le i \le 15$:

\begin{itemize}[label={}]
    \item $(X_1^e,Y_1^e)$, where $X_1^e:=(Par_1^e, Par_2^e)$ and $Y_1^e:=\frac{Par_2^e(\texttt{mag})-Par_{16}^e(\texttt{mag})+1}{2}$,
    \item $(X_2^e,Y_2^e)$, where $X_1^e:=(Par_2^e, Par_3^e)$ and $Y_2^e:=\frac{Par_3^e(\texttt{mag})-Par_{16}^e(\texttt{mag})+1}{2}$,
    \item $\vdots$
    \item $(X_{15}^e,Y_{15}^e)$, where $X_{15}^e:=(Par_{15}^e, Par_{16}^e)$ and $Y_{15}^e:=\frac{Par_{16}^e(\texttt{mag})-Par_{16}^e(\texttt{mag})+1}{2}$.
\end{itemize}

\noindent
Each feature $X_i^e$ consists of parameter values computed at 2 consecutive minutes, whereas each \mbox{label $Y_i^e$} represents the accuracy of the last magnitude estimate in its corresponding feature.

\subsection{Data Scaling}

In order to increase the chances of a successful training outcome, we scale parameter values in the training samples to the range $[0,1]$ by using the \href{https://en.wikipedia.org/wiki/Feature_scaling\string#Rescaling_(min-max_normalization)}{min-max scaling approach} with the Scikit-learn software library~\cite{pedregosa2011}.
This data scaling operation makes data samples more digestible to the training algorithm, which is responsible for teaching the neural network how to predict magnitude accuracies. More precisely, data scaling operations are generally known to improve the effectiveness of gradient descent-based training algorithms, which optimize the weights of neural network models in order to minimize the prediction error by automatically choosing appropriate hyper-parameters (e.g. the learning rate) that are crucial to a fast and optimal convergence towards the final trained network.

\begin{figure}[ht]
\begin{center}
\includegraphics[width=1\textwidth]{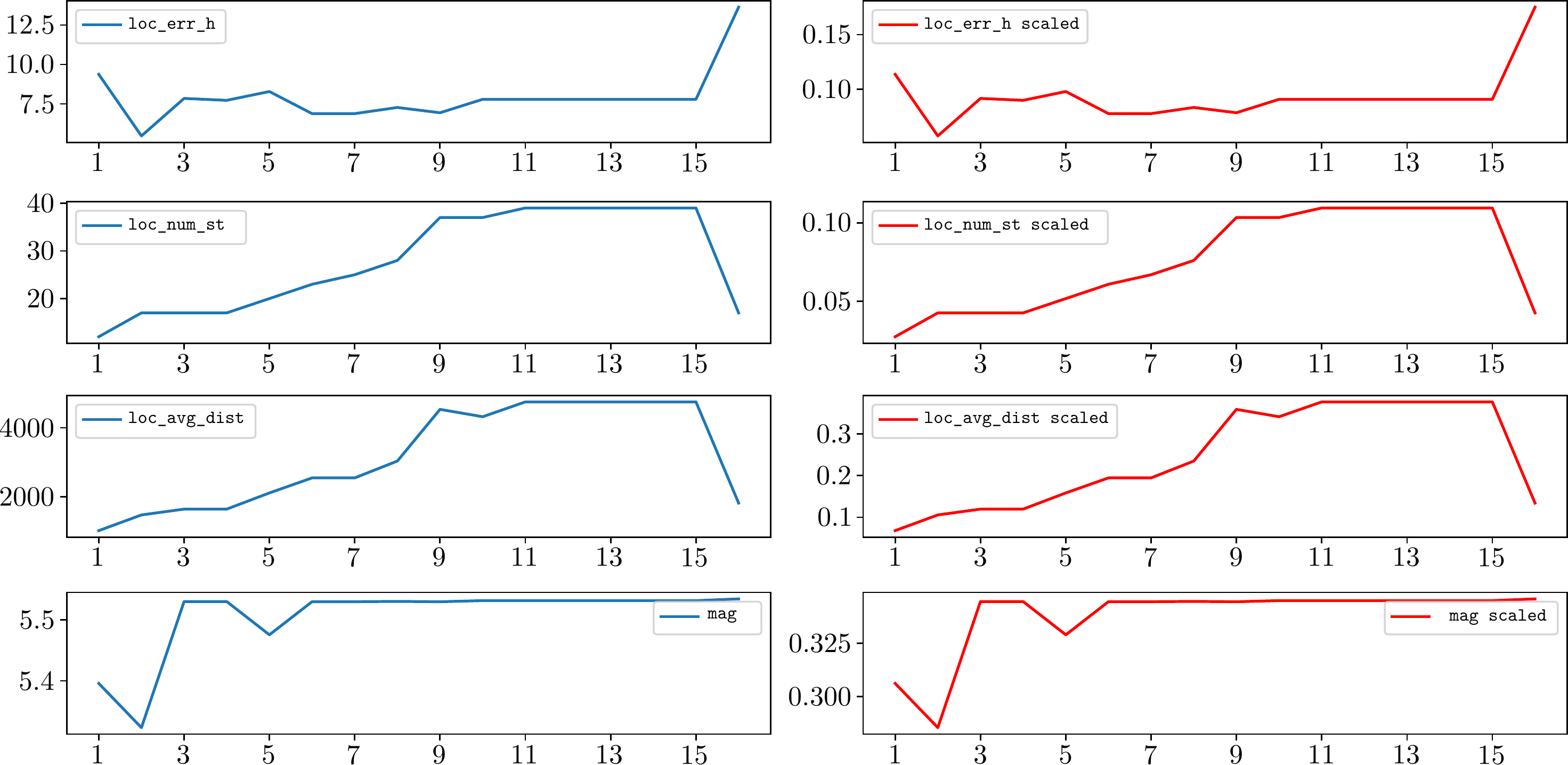}
\caption{\label{figure:scaling}Original data (right) and scaled data (left).}
\end{center}
\end{figure}

\noindent
Figure~\ref{figure:scaling} shows how the trends of parameter values remain unchanged after the scaling process, which means the information about how the parameters develop over time is totally preserved. 

\section{Model Definition and Training}

\noindent
Table~\ref{table:model} illustrates a layer-by-layer definition of our neural network model. First, there are two stacked LSTM layers with rectified activation functions (ReLUs) and input-output dimensions of \mbox{$(K, P)$-$(K, 20)$} and $(K, 20)$-$(10)$, respectively. $K := 2$ is the number of consecutive minutes per input feature, and $P := 12$ is the number of parameter values per minute. Finally, there is a fully connected layer \mbox{(i.e. Dense)} with a sigmoid activation function with input-output dimensions of $(10)$-$(1)$. Both modeling and training processes are performed with Tensorflow and Keras software libraries \cite{abadi2016}\cite{chollet2015}.

\begin{landscape}
\vspace*{15pt}

\renewcommand\arraystretch{2.2}
\begin{table}[h!]
\centering
\begin{tabular}{|p{.18\textwidth}|p{.18\textwidth}|p{.18\textwidth}|p{.18\textwidth}|p{.18\textwidth}|}
\hline
\hline
\textbf{Layer} & \textbf{Type} & \textbf{Act. Func.} & \textbf{Input Dim.} & \textbf{Output Dim.}\\
\hline
\hline
1 & LSTM & ReLU & $(K, P)$ & $(K, 20)$\\
\hline
2 & LSTM & ReLU & $(K, 20)$ & $(10)$\\
\hline
3 & Dense & Sigmoid & $(10)$ & $(1)$\\
\hline
\hline
\end{tabular}
\caption{\label{table:model}Layers of the neural network model.}
\end{table}

\vspace*{5pt}

\renewcommand\arraystretch{2.2}
\begin{table}[h!]
\centering
\begin{tabular}{|p{.35\textwidth}|p{.35\textwidth}|}
\hline
\hline
\textbf{Training Set} & 70\% of the samples\\
\hline
\textbf{Test Set} & 30\% of the samples\\
\hline
\textbf{Optimizer Algorithm} & Adam\\
\hline
\textbf{Loss Function} & Mean squared logarithmic error\\
\hline
\textbf{Batch Size} & 16\\
\hline
\textbf{Validation Set} & 25\% of the training set\\
\hline
\textbf{Number of Epochs} & 100\\
\hline
\hline
\end{tabular}
\caption{\label{table:training}Initial setup of the training process.}
\end{table}
\end{landscape}

\clearpage

\subsection{Training Process}

\begin{figure}[h!]
\begin{center}
\includegraphics[width=1\textwidth]{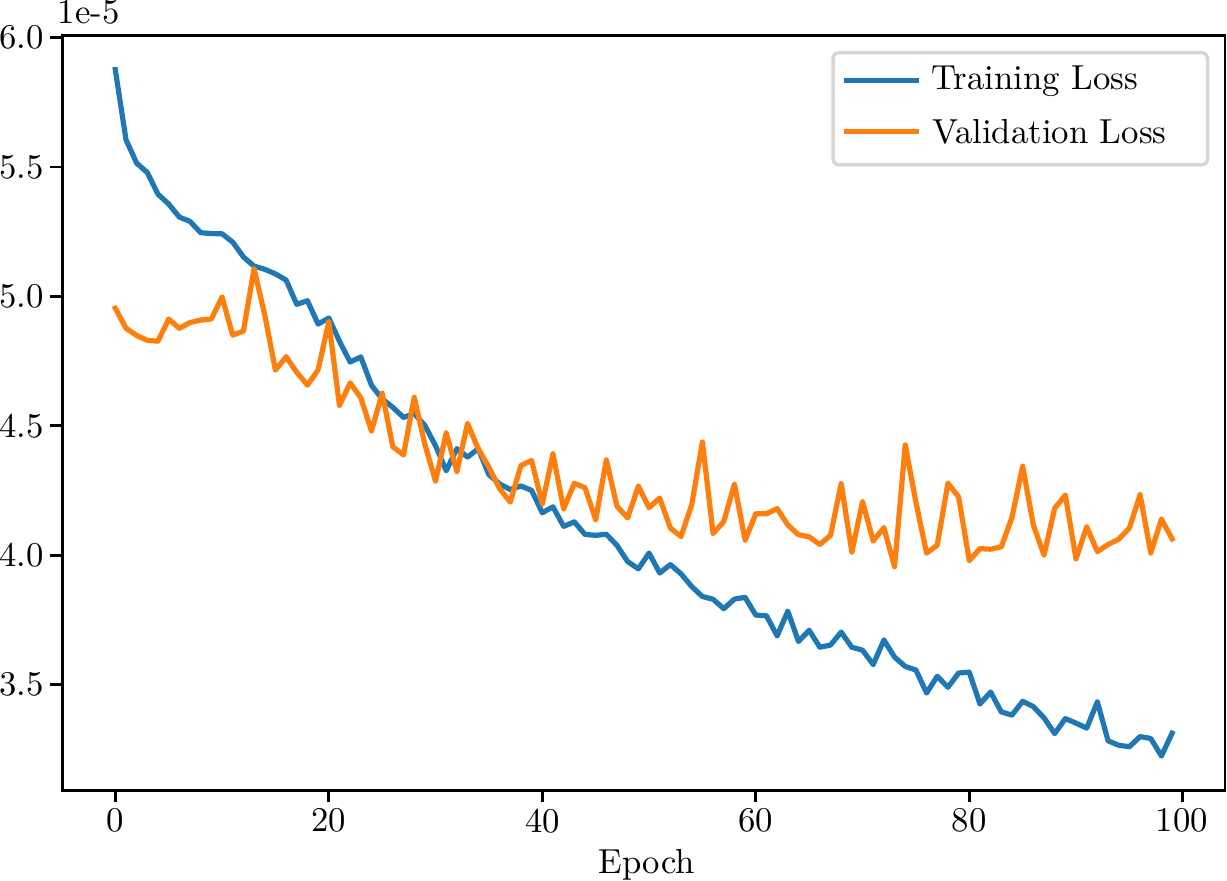}
\caption{\label{figure:history}Training and validation loss during the training process.}
\end{center}
\end{figure}

The initial setup of the training process is presented in Table~\ref{table:training}. The training set consists of 70\% of the samples in the initial dataset shown in Section~\ref{section:samples}, while the remaining 30\% of the samples represent the test set, which is used after the training process in order to evaluate how the final trained model performs on never-before-seen data.
The training algorithm is the Adam optimizer, which iteratively updates the weights of the model in order to reduce the outcomes of the loss function, which consists of the mean squared logarithmic error between the predicted outputs and the actual outputs. The batch size of 16 tells the algorithm how many samples it must process before updating the weights, while the number of epochs of 100 indicates how many times the algorithm must inspect the entire training dataset by randomly drawing batches of samples. At the beginning of each epoch, the algorithm randomly populates a validation set with 25\% of the training set in order to show us a comparison of the training loss and the validation loss in Figure~\ref{figure:history}, which provides us with useful information on how our model behaves during the training process.

\clearpage

\begin{landscape}
\vspace*{40pt}
\begin{figure}[h!]
\begin{center}
\includegraphics[width=1.35\textwidth]{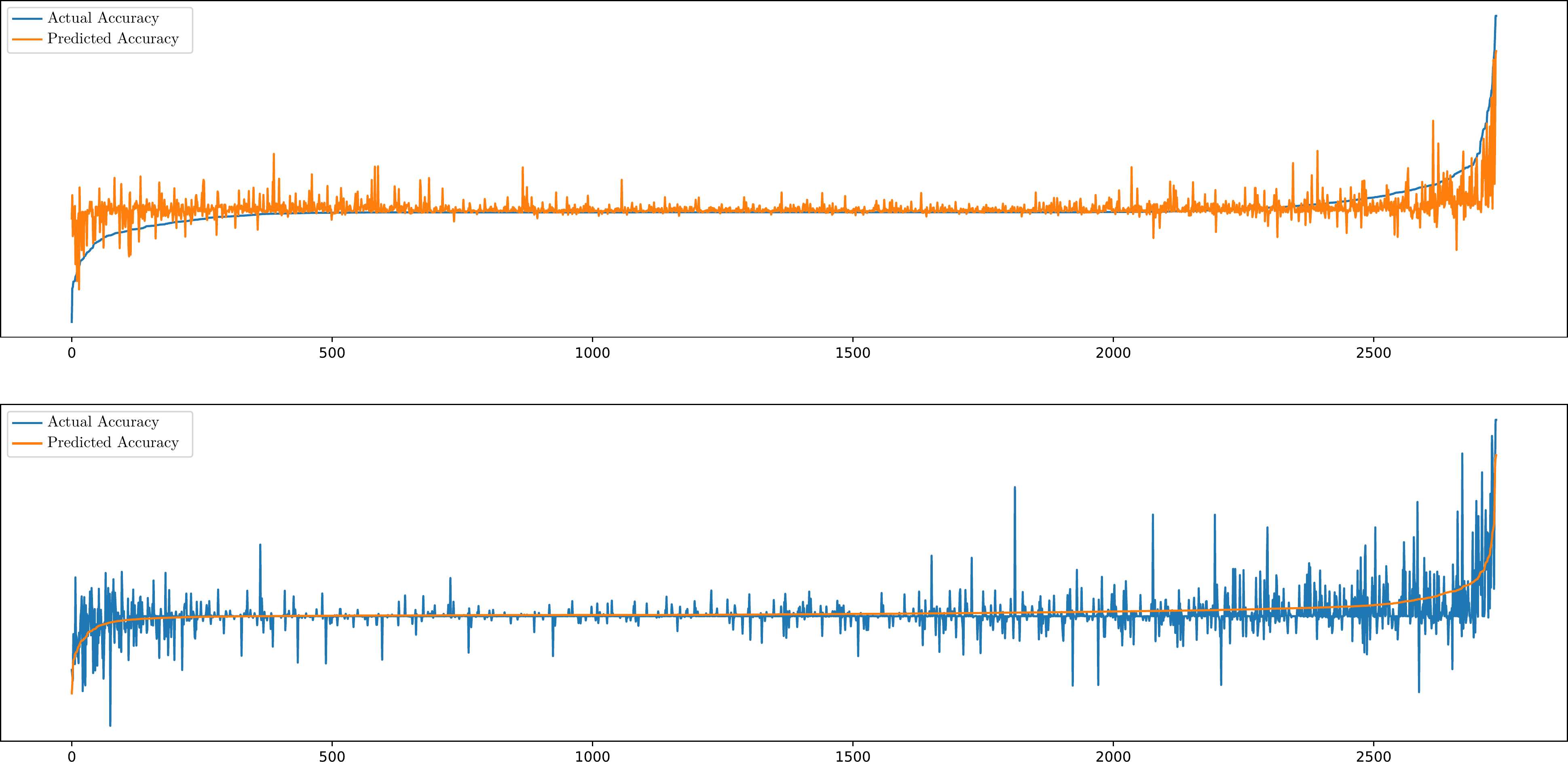}
\caption{\label{figure:test-set-results}Results on the \emph{test set} ordered by actual accuracy (above) and by predicted accuracy (below).}
\end{center}
\end{figure}
\end{landscape}

\section{Evaluation}

Plots in Figure~\ref{figure:test-set-results} show results on the test set by comparing the actual accuracy (in blue) with the accuracy predicted by the model (in orange). The same results are ordered by actual accuracy and by predicted accuracy. In a nutshell, where the orange points coincide with the blue ones that means our trained model was able to perfectly predict the accuracy of the magnitude estimates by Early-est \mbox{(i.e. the actual accuracy)}. These plots provide us with information on how our trained neural network model performs on never-before-seen data.\\

\noindent
Results show the model exhibits the tendency to correctly predict errors whenever the magnitudes are under- or overestimated (left-hand and right-hand side in the plots, respectively).

\section{Conclusion}

In spite of the limited amount of historical data in this initial analysis, results show the designed network exhibits the tendency to correctly predict errors whenever the magnitudes are under- or overestimated. This fact indicates that the historical data are suitable for the automatic training of the neural network presented in this report.

\section*{Acknowledgments}

This work would not have been possible without the help of Dr. Franco Mele, Dr. Stefano Lorito and the \href{https://www.ingv.it/cat/en}{Tsunami Alert Center} group at \href{https://www.ingv.it}{INGV}.

\end{document}